\documentclass[prd,eqsecnum,noshowpacs,nofootinbib,notitlepage]{revtex4-1}
\usepackage{amsmath,amssymb,amsthm,amsbsy}
\usepackage{amsfonts}
\usepackage{comment}
\usepackage{color}
\usepackage{graphicx}
\usepackage{mathptmx}


\newtheorem{theorem}{Theorem}

\begin{document}
\title{Causal concept for black hole shadows}

\author{Masaru Siino}
\email{msiino@th.phys.titech.ac.jp}
\affiliation{Department of Physics, Tokyo Institute of Technology, Tokyo 152-8551, Japan}

%\author{Haruki Terauchi}
%\email{terauchi@th.phys.titech.ac.jp}
%\affiliation{Department of Physics, Tokyo Institute of Technology, Tokyo 152-8551, Japan}

%\author{Daisuke Yoshida}
%\email{d.yoshida@physics.mcgill.ca}
%\affiliation{Department of Physics, McGill University, Montr\'eal, QC, H3A 2T8, Canada}

 \begin{abstract}
Causal concept for the general black hole shadow is investigated, instead of the photon sphere.
  We define several `wandering null geodesics' as complete null geodesics accompanied by repetitive conjugate points, which would correspond to null geodesics on the photon sphere in Schwarzschild spacetime. We also define a `wandering set', that is, a set of totally wandering null geodesics as a counterpart of the photon sphere, and moreover, a truncated wandering null geodesic to symbolically discuss its formation.
 Then we examine the existence of a wandering null geodesic in general black hole spacetimes mainly in terms of Weyl focusing.
 We will see the essence of the black hole shadow is not the stationary cycling of the photon orbits which is the concept only available in a stationary spacetime, but their accumulation.
A wandering null geodesic implies that this accumulation will be occur somewhere in an asymptotically flat spacetime.
% A generalization of the photon sphere for general black hole spacetime based on the causal structure is proposed and named wandering set.
%To generalize the concept of the photon sphere, wandering set of wandering null geodesics is defined.
% We demonstrate the existence of the wandering set in general black hole spacetimes, under the global hyperbolicity. Furthermore, we see that the wandering null geodesics destined to be accompanied by repetitive conjugate points, which is consistent with the fact that the null geodesics do not escape to the future null infinity.
 \end{abstract}

\maketitle
\section{Introduction}

Recently black hole physics has been getting the two great advances in the observation for astrophysical phenomena around the black hole.
The first observed gravitational waves~\cite{Abbott:2016blz} are consistent with those generated by a merger of a black hole binary in general relativity.
This is the first, but indirect evidence that a black hole merger would occur in our universe. Thus, it is still important to develop approaches to observe such dynamical black holes as mergers of black holes more directly.

One of the approaches would be optical observation. 
Indeed, the other very recent advance, that is, the real image of photons around the massive black hole at the center of M87 is excitingly reported by the Event Horizon Telescope (EHT) -- a planet scale array of eight ground based radio telescopes forged through international collaboration \cite{Akiyama:2019cqa,Akiyama:2019brx,Akiyama:2019sww,Akiyama:2019bqs,Akiyama:2019fyp,Akiyama:2019eap}.
Of course, by definition of the black hole we cannot see its event horizon directly.
Black hole region is defined as the complement of the causal past of future null infinity, and the event horizon is defined as the boundary of the black hole region. The dynamics of the event horizon directly reflects that of black holes itself. Especially, the typical situation for gravitational radiation like black hole coalescence accompanied by the change of topology of the event horizon. In this sense, the topological notion for binary black holes has been developed to investigate its topological appearance so far~\cite{Siino:1997ix,Siino:1998dq,Husa:1999nm,Cohen:2011cf,Bohn:2016soe,Emparan:2017vyp}. 
We cannot see, however, any event horizon optically because the event horizon is a null surface generated by null geodesics without future endpoints~\cite{Hawking:1973uf}, and hence our (asymptotic observer's) causal past does not intersect the event horizon.
Instead, the form of black holes would be seen as the shadows~\cite{Falcke:1999pj,Broderick:2009ph,Broderick:2010kx} rather than the event horizon even by the optical observation.

Actually shadows of various spacetimes have been studied so far\cite{Bambi:2010hf, Bambi2019,Nitta2011,Yumoto:2012kz,Abdujabbarov:2012bn,Grenzebach:2014fha}, which have also been studied in modified theories of gravity\cite{Amarilla:2010zq,Amarilla:2011fx,Dastan:2016vhb,Cunha:2016wzk,Vagnozzi2019}.  
Black hole shadows are well understood in the context of the photon sphere\cite{photonsphere,Hod2013,Hod2018} in static spherically symmetric spacetimes. A photon sphere is the set of the unstable circular photon orbits, and there is a certain discontinuity around a photon sphere\cite{Sanchez1978,Decanini2010}.
For example, in Schwarzschild spacetime the photon sphere is $r=3M$ sphere, and photon orbits which immediately fall into the black hole or which go away to the infinity are accumulated around it\cite{Virbhadra2000}.
Consequently, for an asymptotic observer the photon sphere will be bright if there are sufficient light sources outside of the sphere, while the inside of it will be dark.

Now we can expect that we would simultaneously observe an information from the same astronomical object both by gravitational wave and the black hole shadow in the near future.
If so, the astronomical object will be including a dynamical black hole.
The theoretical progress, however, has been just short in general situations of the black hole shadows, which relies on the study of the photon sphere for the general black hole.

Moreover, even for understanding the observation\cite{Akiyama:2019cqa,Akiyama:2019brx,Akiyama:2019sww,Akiyama:2019bqs,Akiyama:2019fyp,Akiyama:2019eap}, though it may be sufficient to analyze static photon sphere in order to understand the bright region, it will require understanding the formation of the photon sphere to judge what is indicated by the dark region. The past directed null geodesics approaching the event horizon have reached it far in the past as long as it is after the formation of the event horizon.
For example, in a dynamically formed black hole, a past directed null geodesic oriented to the direction of the central dark region in the field of view can never reach to the event horizon by definition, in any far in the past. 
Rather, while this past going null geodesics approach the event horizon up to the formation era, they may reach any source of photons without concerning the event horizon before the formation era.

A concept of a photon sphere is generalized to a photon region\cite{teo2003spherical, Grenzebach:2014fha} for stationary black holes, but it is not for general black hole spacetimes.
%{a} %one of the
Another concept toward generalization of the photon sphere for general black holes may be a photon surface~\cite{Claudel:2000yi}.
In static spherically symmetric situations, however, photon surfaces include not only the photon sphere, but also general null surfaces even including the event horizon.  Therefore, it is doubtful whether generally there is the accumulation of the photon orbits around the photon surface. 
Then we have a question whether the photon sphere can play an important role for more general spacetimes.
Especially, we are interested in that on highly dynamical situations, for example the formation of black holes or collision of black holes, if we would like to relate them to gravitational wave radiation.

\begin{figure}[hbtp]
\begin{center}
 \includegraphics[height=12cm,clip]{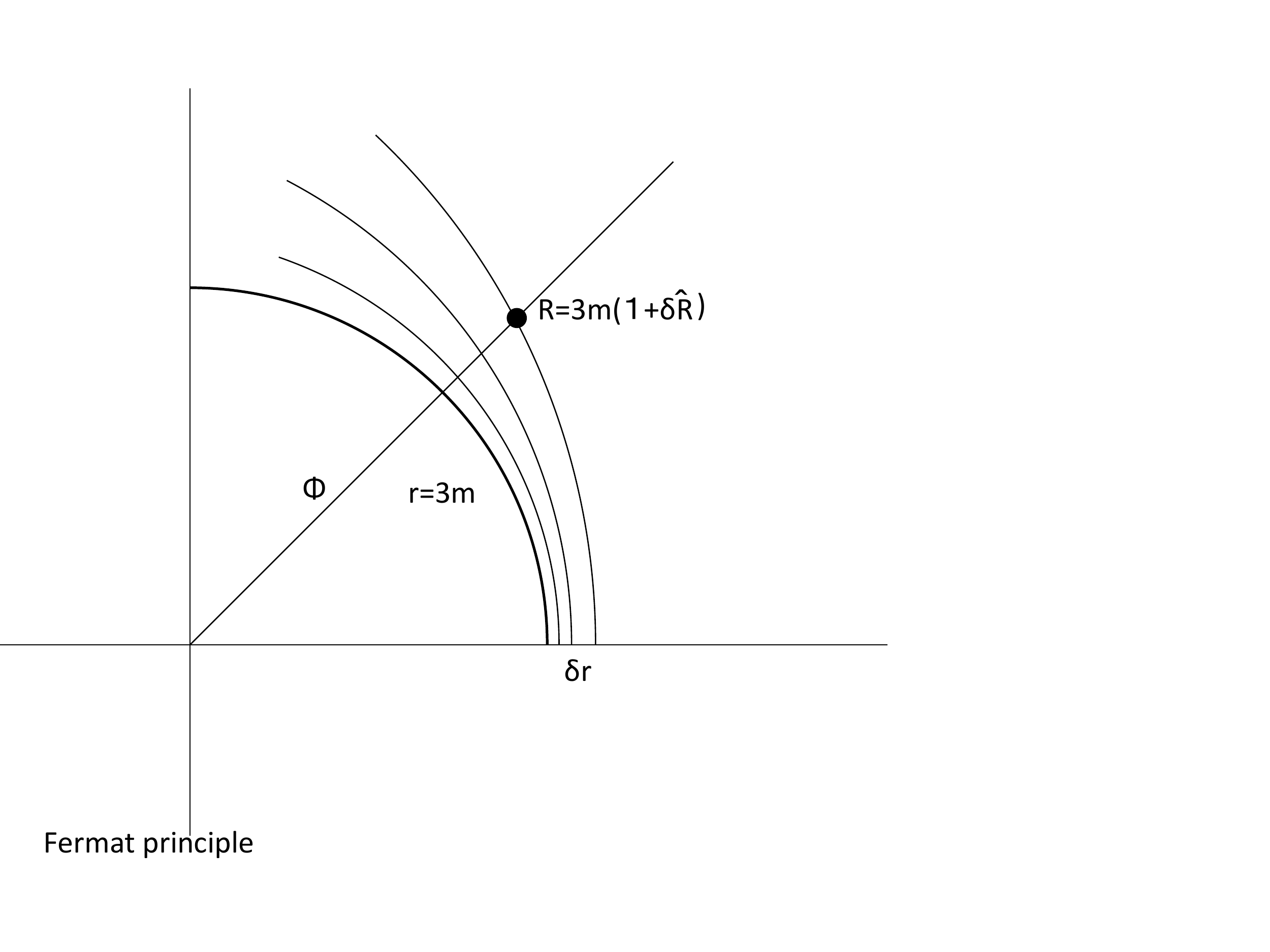}
 \label{fig1}
 \end{center}
 \caption{A figure of the null congruence near the region $r=3m$ on the equatorial plane on the Schwarzschild spacetime plotted in $(r,\phi)$ plane: the most inner null geodesic in this figure represents the unstable circular orbit on $r=3m$. Our interest is where the most outer null geodesic starts.  
 \label{fig:1}
 }
   \end{figure}

To guess the possibility of similar phenomena to the accumulation in more general situations, we simply discuss structural stability of such an aspect that photon orbits accumulated around the unstable photon orbits in the Schwarzschild spacetime. Let $\{t, r, \theta, \phi\}$ be Schwarzschild coordinates and let $m$ be the mass of the black hole.
In a static universe, Fermat's principle determines the orbit of a light ray as the stationary point of the time function: The Schwarzschild time, which takes for a light ray to get to a goal point from a starting point.
Here we consider a light ray on an orbit $r(\phi)$ from $ (r, \theta, \phi)=(r(0),\pi/2, 0) $ to $(r(\Phi)=R,\pi/2,\Phi)$ near the sphere $r=3m$ on the equatorial plane $\theta=\pi/2$ (see fig\ref{fig:1}).
Introducing small variables $\delta \hat{R}=R/(3m)-1$ and $\delta \hat{r}= r(0)/(3m)-1$, the time function of this light ray can be obtained
\footnote{
In (\ref{timefunction}), we write the leading for $\Phi$ of order $\delta {\hat{r}}^{0}$, $\Phi^{2}/3 $, though this term is included in ${\cal O}(\delta \hat{R}^2, \Phi^{1}, \delta \hat{r}^4)$. This is because the form of the potential function is important in catastrophe theory.
}
 as
\begin{align}
\frac{t[\delta \hat{R},\Phi](\delta \hat{r})}{9\sqrt{3}m/\Phi}&
=\left(-1 + \frac{3}{\Phi ^2}  \delta
   \hat{R}\right) \delta \hat{r}^3
+\left(
    \frac{1}{2}+ \delta \hat{R} \right) \delta \hat{r}^2
- \delta \hat{R}\delta \hat{r}
+\frac{\Phi^{2}}{3}
+{\cal O}(\delta \hat{R}^2, \Phi^{1}, \delta \hat{r}^4).
   \label{timefunction}
\end{align}
This function is reduced to $t= x^3+ax+c+{\cal O}[x^4]$ by a diffeomorphism in variable space $(\delta \hat{R},\Phi,\delta \hat{r})$ around the origin.
In this case, the Fermat's principle says, considering a photon reaching a point $ (r= R>3m,\phi=\Phi)$, it should be start the point on the wave front $\phi=0$ where $\partial t[\delta \hat{R},\Phi](\delta \hat{r})/\partial \delta \hat{r} = 0$. In the context of catastrophe theory, 
$\delta \hat{R}$ and $\Phi$ are control variables and $\delta \hat{r}$ is a state variable.  The accumulation of the orbits is realized by this time function, which is called fold-catastrophe~\cite{poston2014catastrophe}. 
Furthermore, the Thom's theorem states such a catastrophe set is structurally stable.  
The structural stability means, for example, the stability under small changes of the geodesic equation. When such perturbation occurs dynamically the orbits of light rays will shift, but the accumulation of the orbits will not be broken.\footnote{Here we only consider orbits in an equatorial surface and also perturbation is limited to spherically symmetric one.  If we would like to consider non-spherical perturbation, we should impose the reflection symmetry to the perturbation to maintain this aspect.} From such structural stability, we expect similar accumulation and discontinuity will be present also in dynamical black hole spacetimes as long as they are adiabatically slow.

The purpose of the present article is to develop and discuss causal concept for the black hole shadow instead of such unstable circular orbits of photon, for general black hole spacetimes. Especially, we would like to discuss whether anisotropic or highly dynamical black holes allow us to introduce this concept, while the authors develop it in terms of any coordinate system according to the spherical symmetry\cite{Schneider2018,Cao:2019vlu,Mishra2019,Bisnovatyi-Kogan2018} or resembling stationary coordinate\cite{Yoshino2020,Mars2017}.

This paper is organized as follows.
In the next section, we develop causal concept possibly corresponding to the photon sphere in Schwarzschild spacetime, which is named wandering set, and demonstrate it is closely related to the existence of the conjugate point.
The third section provides analysis for the existence and the formation of such a geometrical structure, considering the truncation of the wandering null geodesics.
Furthermore, we will discuss the relevance of the wandering null geodesic for the observation of the general black hole shadow in the fourth section.
The final section is devoted to summary and discussions.

Especially, the definition of null geodesic congruence and its conjugate point obey the Wald's text book\cite{Wald:1984rg}.
In the present work, we treat only inextendible geodesics. Moreover, the spacetime is considered to be globally hyperbolic without any caution.

\section{definition of wandering null geodesic}
A direct idea to consider the photon sphere in general spacetime will be to find unstable cycling photon orbits from the form of a potential function included in a null geodesic equation. 
Nevertheless, in considering general situations the geodesic equation will not be realized by the potential function because of the evolution of the black hole.
Another idea might be the photon surface~\cite{Claudel:2000yi} as mentioned above. The photon surface, however, possesses an unfavorable characteristic: it exists in Minkowski spacetime and does not exist in Kerr black hole spacetime.
Moreover, since the definition of the photon surface is intrinsic, it does not seem to be useful to find the accumulation of the orbits.
As our aim is to consider concept resembling a photon sphere, and the phenomena\cite{Mach2013,Chaverra2015,Cvetic2016,Koga2016,Koga2018,Koga2019} accompanying it in various dynamical black hole spacetimes, we incline to think of their causal concept, rather than the analytical terminology of null geodesic equation, while some ideas for notion of trapped geodesics has been just analytically investigated\cite{Gibbons2016,Cunha2017,Galtsov2019,Galtsov2019a,Koga2019a,Yoshino2020,Mars2017}.

Considering an asymptotically flat spacetime~\cite{Hawking:1973uf,Wald:1984rg}, inextendible complete curves go toward   the boundary of the spacetime manifold composed of future (past) null infinity $\cal I^+,\cal I^-$, future (past) timelike infinity $i^+, i^-$ and spatial infinity $i^0$.
The key observation here is that there are null geodesics which go toward future {`timelike'} infinity $i^+$.  
For example, in the case of Schwarzschild spacetime, the null geodesics of unstable circular photon orbits on $r=3m$ go to timelike infinity $i^+$.
Is this feature common, among the phenomena which should occur with the accumulation of photon orbits around black holes?

Though such a geodesic will not be allowed in asymptotically simple spacetimes~\cite{penrose1986spinors}, general black hole spacetimes may have such null geodesics under the condition of weak asymptotic simplicity.
Though it is not clear whether the accumulation of photon orbits exist around a null geodesic going to timelike infinity, we expect that such a null geodesic characterizes the black hole spacetime even in dynamical situations.

 In the sense of conformal completion, however, the unphysical manifold $\overline{M}$ does not include the point of timelike infinity $i^+$, since the conformal boundary is not smooth there.
To include $i^+$ in the boundary of the manifold, one may consider c-boundary~\cite{Geroch:1972un}. Then we see the boundary will have Hausdorff topology, but it is not easy to handle it.
Therefore, it would not be good choice to define the concept that geodesics go to $i^+$ in terms of terminal indecomposable past. 
Consequently, we will rather concentrate on null geodesics which do not fall into a black hole and do not go away to future null infinity.
We may call such a null geodesic a `neutral' null geodesic $\gamma_n$ as a counter part of the circular photon orbits.
Nevertheless, we see that the concept of neutral null geodesics is including a little different null geodesic from the photon sphere, since the event horizon is generated by such neutral geodesics.

In the rest of this section, we enter into the discussion using null geodesic congruence. Our concept that the null geodesic does not go away and not fall into the black hole will be realized by a `wandering' characteristic of null geodesic congruence in the following of this section, which should be complete and inextendible\footnote{In the present work, we only consider inextendible one since the parameter of the null geodesic for light ray is to be set to affine parameter.}.
Then we shall consider a wandering null geodesic $\gamma_w$ and the wandering set $W$, which is the set of $\gamma_w$, to construct causal concept for a general black hole shadow.

Now we investigate the geometrical aspects for a wandering null geodesic.
To analyze behaviors of light rays accompanying the wandering null geodesics, we will discuss a null geodesic congruence around the wandering null geodesic.
Each component of the deviation equation~\cite{Wald:1984rg} for such a congruence of null geodesics with tangent vector $k^a$ is given by
\begin{align}
\frac{d\theta}{d s}=-\frac12\theta^2-\hat{\sigma}_{ab}\hat{\sigma}^{ab}-R_{cd}k^ck^d \label{eq:dev1}\\
k^c\nabla_c\hat{\sigma}_{ab}=-\theta\hat{\sigma}_{ab} +\widehat{C_{cbad}k^ck^d}, \label{eq:dev2}
\end{align}
where $s$ is the affine parameter of null geodesics, $\theta$ and $\hat{\sigma}_{ab}$ are the expansion and the shear of the null congruence, respectively. Here `` $\ \hat{}\ $ '' represents an element of the equivalence class with respect to the equivalence relation $x^a$ and $y^a$ is defined as equivalent if $x^a -y^a \propto k^a $ (See Ref. \cite{Wald:1984rg}).
Supposing that the wandering null geodesic does not have conjugate points at all, $\theta$ should be positive for the infinite range of the affine parameter.
Then, if we can omit the curvature term, the equation has a solution in the form of $\theta(s)=\Theta/(s-s_0), \sigma_{ab}(s)=\Sigma_{ab}/(s-s_0)$, and it is a solution for a straight light in a flat spacetime.
Nevertheless, since the congruence does not go to infinity, the curvature terms must be non-zero under the null generic condition: $k_{[e}C_{a]bc[d}k_{f]}k^ek^c\neq 0$. In time the curvature terms should dominate the equations since $\theta,\hat{\sigma}_{ab}$ are dumping in $\sim1/(s-s_0)$.
Then $\hat{\sigma}_{ab}$ becomes large and the first equation is dominated by the $\hat{\sigma}$-term.
Therefore, soon $\theta$ becomes negative $\theta_0$ and then they will have a conjugate point since $\theta$ will diverge within affine length $s\leq 2/|\theta_0|$.

 \begin{figure}[hbtp]
  \begin{center}
\includegraphics[height=12cm]{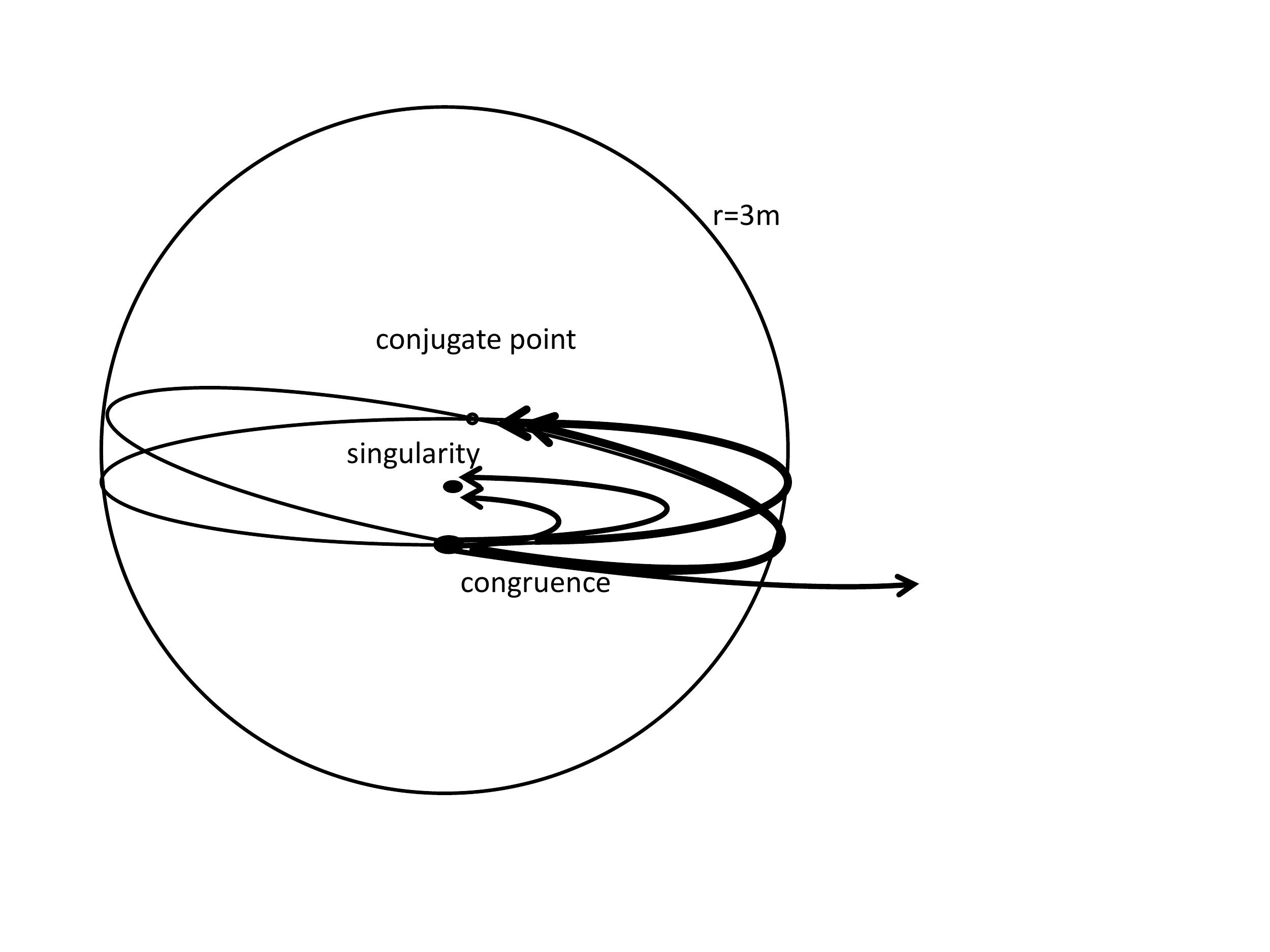}
 \caption{
An image of a null congruence emanating from the south pole on $r=3m$ on the Schwarzschild spacetime. The sphere represents $r=3m$ sphere.  A null geodesic outer directed than the wandering null geodesic goes to future null infinity, and a null geodesic inner directed goes to the singularity. 
However the wandering null geodesic, which is on $r=3m$ sphere, passes through conjugate points repetitively.}
   \label{fig:3}
  \end{center}
 \end{figure}

As we are considering a globally hyperbolic spacetime, the existence of conjugate points imply such a null geodesic come into the inside of the chronological future of the starting point.
Redefining the null geodesic congruence at the conjugate point again, the same logic suggests that along this null geodesic conjugate points repetitively appear. 

This situation can be easily illustrated in the case of Schwarzschild spacetime (see Fig. \ref{fig:3}).
Considering a family of null geodesics starting from the south pole of the photon sphere around $\gamma_{c}$ corresponding to an unstable circular photon orbit, from the above discussion $\gamma_c$ has conjugate points repetitively. Since the photon sphere is a sphere, it is clear that the null geodesics on the sphere starting from the pole pass through conjugate points on the north and south poles repetitively.

 At the starting pole, the null geodesics of the congruence, going to the inner side of the photon sphere immediately fall into the black hole.
On the other hand, the null geodesics of the congruence, going to the outside of the photon sphere will go to the region where the curvature terms can be omitted.
It will continue going toward infinity since $\theta\propto d (\ln A) /ds\sim 1/(s-s_0) $ is integrated as $A\sim(s-s_0)$, where the area $A$ of the section of the congruence is divergent.
There, the null geodesic will not have conjugate points anymore and go to null infinity since it becomes a null generator of the boundary of any future sets.
In those situations, not only $\theta$ but also $\sigma_{ab}$ may diverge at the conjugate point.

  \begin{figure}[hbtp]
  \begin{center}
\includegraphics[height=12cm]{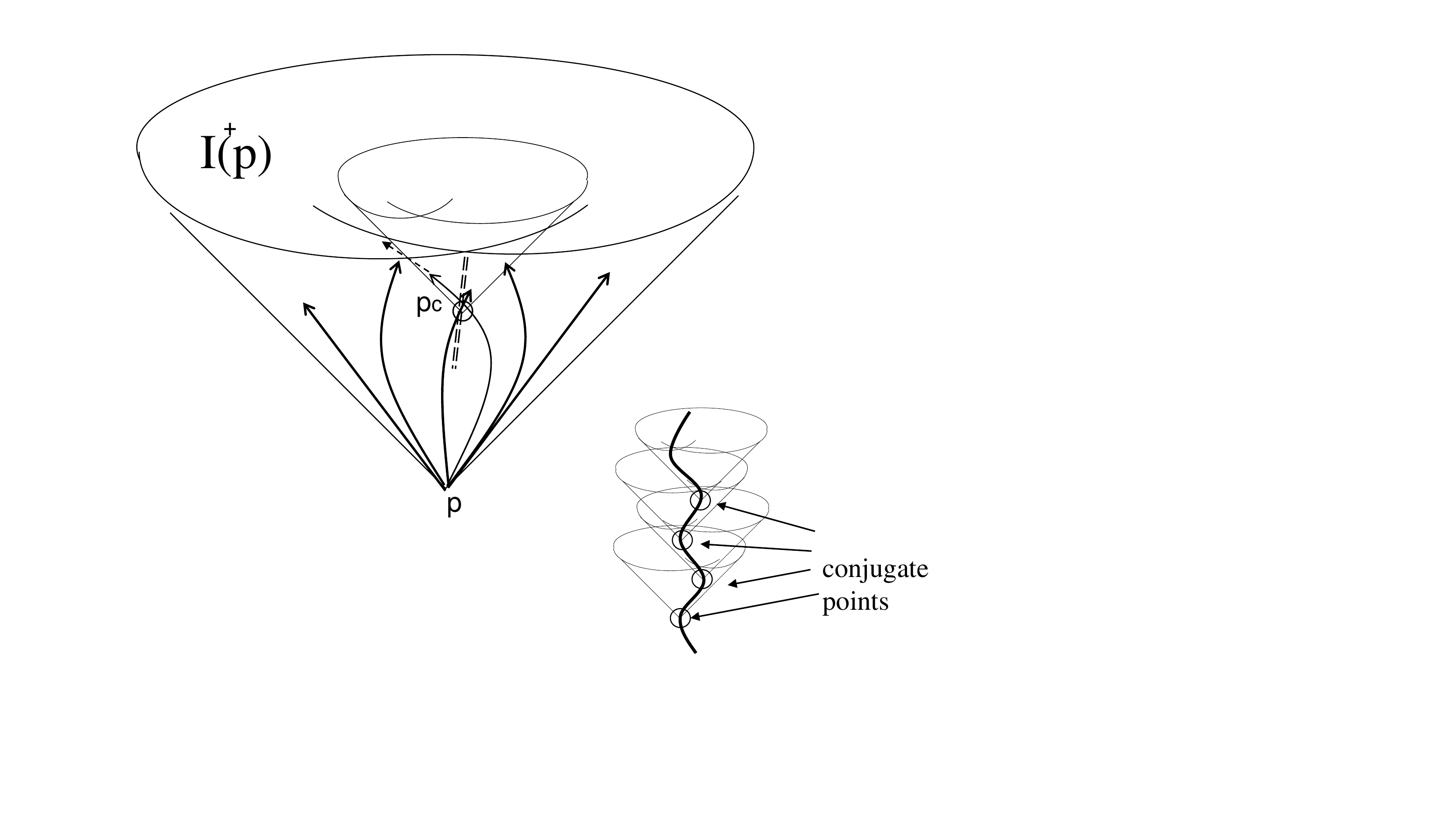}
 \caption{Through the conjugate point $p_c$, a null geodesic invades into ${\rm Int}(I^+(p))$, and $p\sim p_c$ is generators of a $\partial(I^+(p))$.
Beyond $p_c$, $p_c\sim ...$ is not the generator of $\partial(I^+(p))$ but becomes a generator of the boundary of another future set $I^+(p_c)$ till a next conjugate point appears.
Repeating this argument, conjugate points turn out to appear repetitively on this null geodesic. }
   \label{fig:4}
  \end{center}
 \end{figure}

In a sense, the conjugate point realizes that the null geodesic invades into the chronological future of the starting point under the global hyperbolicity (see Fig. \ref{fig:4}). 
This situation is of small variation for the null geodesic generator of the chronological future\cite{Hawking:1973uf}\cite{Wald:1984rg}.
Roughly speaking, the existence of conjugate points means the null geodesics pass round and round around the black hole, repetitively visiting the conjugate points. 
Of course, one may argue that the conjugate point is not necessarily required for the null geodesic to invade into the chronological future of the starting point (topological non-triviality may globally cause it, for example in a static spacetime of flat $S_1\times R^2$ spatial section).  As illustrated in the case of Schwarzschild spacetime, however, the repetitive occurrence of the conjugate points is corresponding to going around the event horizon. 
%Our only concern is the case of the conjugate points. 

Here we should notice that the wandering null geodesic excludes the null geodesic which is a generator of the event horizon.
Since the horizon generator has no future endpoint\cite{Hawking:1973uf} under the global hyperbolicity, any conjugate point is not allowed. Then the event horizon generators are not wandering instead of neutrality.
This is comfortable to discuss the observational relevance of the wandering set.

We insist that the usual wandering of the null geodesic around the black hole region should be related to the conjugate point of complete null geodesics.
Then we will just define a `wandering null geodesic' $\gamma_w$, which is complete, inextendible, and with unlimited number of conjugate points.

Supplementary, we also define
% the future (past) wandering set $W^+$ ($W^-$) by the set of 
a future (past) wandering null geodesic $\gamma_w^+$ ($\gamma_w^-$) from $p$, which is a future (past) -directed complete and inextendible null geodesic with unlimited number of conjugate points starting from a point $p$.
Then we define also a totally wandering null geodesic $\gamma_{wT}$ which is past and future complete and possesses unlimited number of repetitive conjugate points in both future and past directions. The totally wandering null geodesics will coincide with the circular photon orbit of the Schwarzschild spacetime.

Thus, thinking the `wandering set' of the totally wandering null geodesics, we expect they give causal concept for the black hole shadows in general spacetime instead of the photon sphere.
It should be noticed, however, that generally a wandering null geodesic has any uneven distribution of conjugate points.
In a single future wandering null geodesic, for example, we would distinguish a null segment approaching the black hole, from a null segment going around the event horizon with the occurrences of the conjugate points.
In the case of Schwarzschild spacetime these aspects enable us to have future (past) wandering null geodesics from all the points outside of the event horizon.
Since that means a set of them is corresponding to the whole of the outside of the black hole, such a set does not make sense.
Rather we will define a wandering set as a set of the totally wandering null geodesics as $W=\{\gamma_{wT}\}$.

Here it should be noted that the totally wandering null geodesic and the wandering set will only exist in eternal black hole spacetime where the black hole not created and never die\footnote{Strictly speaking, the wandering null geodesic is possible also without black hole. Here we omit it since it would be difficult because of the reality of matter field.}.
Nevertheless, not only in stationary black hole spacetime, they would exist but also in eternally evolving black hole spacetime and merging eternal black hole spacetime.

On the other side, when we consider the formation of the black hole shadow in dynamical black hole, we had better concentrate on the future wandering null geodesics, though its set possesses above mentioned unfavorable nature.
  In the next section, we shall consider another wandering null geodesic by truncation of the future wandering null geodesic.
That would be called a truncated wandering null geodesic.

Here we summary these derivative definitions of the wandering null geodesics and wandering set, related to the possible aims in the study of the black hole shadow in general situations.

\begin{itemize}
\item 
To investigate the counterpart of the circular orbit for stationary black hole, eternally evolving black hole or merging eternal black hole we will think of totally wandering null geodesic, which possesses an infinite number of conjugate points in both future and past directions.
Hence, for example, the wandering set will coincide with the photon sphere in Schwarzschild black hole. Since the photon sphere, however, is the set of the ``unstable'' circular orbit, the wandering set does not always coincide with the photon sphere even in static spherical black hole.

\item To see that the wandering null geodesic gives causal concept for the black hole shadow, we think of the totally wandering null geodesic for a globally hyperbolic spacetime in the past direction.
Though observational detail should be generally discussed incorporating also future and past wandering null geodesics\footnote{It would be discussed more carefully in our forthcoming work\cite{GG3}.}, in fourth section, for simplicity we investigate it as the null geodesic congruence around the totally wandering null geodesic. 

\item In order that we symbolically discuss the formation of the wandering null geodesics, and compare it with the black hole formation, we will truncate the future wandering null geodesic at the first conjugate point, artificially in the next section. 
We call this derivative wandering null geodesic as a truncated wandering null geodesic.
Of course, we expect that it will not cause so great change to the formation of this truncated null geodesic though this process includes small ambiguity depending on the starting point of the null geodesic or the initial condition of the deviation vector.
For that reason, we note the truncated null geodesics will not form thin shell even in the Schwarzschild spacetime.
\end{itemize}

%Now we will speculate the composition of the null geodesic congruence $\{\gamma\}$ around  
%such a wandering null geodesic $\gamma_w$.
%If we consider a pretty long null geodesic $\gamma'_w$ with large but finite affine length, 
%, a number of $\gamma'_w$ would be contained in $\{\gamma\}$ and a sub-congruence of $\{\gamma'_w\}$ will be with finite measure' in $\{\gamma\}$. If we enlarge the affine length $k_m$, the `measure' of the sub-congruence will decrease.  
%Intuitively, in the limit of $k_m\rightarrow \infty$, the sub-congruence $\gamma'_w$ will
%become measure zeroes'. In the congruence $\{\gamma\}$ at the limit, it is not so clear whether the $\gamma_w$ is isolated or not though there might be several $\gamma_w$s.
%Can we expect under any appropriate conditions that $\{\gamma_w\}$ is a surface
%dividing the congruence into ingoing one and outgoing one as in the case of Schwarzschild $spacetime?????? Such aspects shall be studied in generic situations in the context of catastrophe theory.

%Here it should be noted that the congruence does not mean the existence of real such a family of wandering null geodesics, since the Jacobi field is only a first order differentiation of null geodesics. The higher-order differentials may prevent such an infinitesimal one
% parameter conjugate family to be integrated to a finite scale, while in the case of the Schwarzschild black hole, the spherical symmetry insures that the higher order differentials disappear and the one parameter family generates the real photon sphere. 

\section{truncated wandering null geodesic and the formation of black hole}
Here we want to discuss the formation and the existence of the wandering null geodesics.
Since there will be no such conjugate points in past of the dynamically collapsing black hole spacetime, future wandering null geodesic which is complete future null geodesic accompanied by repetitive conjugate points, should be considered.
Since generally there exist the future wandering null geodesics starting from a point in wide regions of black hole spacetime far from the black hole region, for example, in the Schwarzschild black hole, from all the points outside of the event horizon, there is a null geodesic which is approaching to the unstable circular orbit of a photon and converging to it, which means that the null geodesic is accompanied by the permanently repetitive conjugate points.
And then the set of the future wandering null geodesics will cover all of the outside of the event horizon.
This generality of the wandering null geodesic, however, is in a sense terrible thing that we cannot judge when the photon sphere phenomenon started.
%Furthermore it has been noticed that the wandering null geodesic on the unstable orbit of the Schwarzschild black hole destined to possess conjugate points repetitively.

We think of the future wandering null geodesic starting from an arbitral point has two phases of segments. 
As suggested at the bottom of the previous section, one is a wandering phase in which the wandering null geodesic stays and wanders in a chronal region with the repetitive conjugate points. The other is an approaching phase where the null geodesic is going to and approaching   a wandering orbit near the event horizon without a conjugate point.
Then it is natural to define a `truncated future wandering null geodesic' for the wandering phase by cutting it off at the first conjugate point, and discarding the segment from the starting point to the first conjugate point in the approaching phase.
%Furthermore, we define a `truncated future wandering set' as the set of the truncated future wandering null geodesics.

Under the global hyperbolicity, since the conjugate point is sufficient for a null geodesic to stay a chronal region, we have regarded the conjugate point as the mark of wandering of a null geodesic.
On the other hand, the truncated wandering null geodesic is not defined so that it forms a thin shell like a photon sphere, since the truncation includes small ambiguity depending on the starting point of the null geodesic or the initial condition of the deviation vector. 
Then the existence of the truncated wandering null geodesic possesses only symbolic meaning about the wandering phenomenon of null geodesics and we do not define the set of the truncated wandering null geodesic.
Consequently, we will use the future truncated wandering null geodesics in order to symbolically discuss the existence of cycling null geodesics, which are segments of the future wandering null geodesic after the first conjugate points.
To identify the formation of structure like photon spheres, we will study the occurrence of the conjugate point in the region where a black hole is in formed.

 In the discussion of the conjugate point, we will see that conjugate points will be formed due to the essential contribution of the Riemann curvature for the deviation equation.
Thinking it intuitively, we will suspect that such a geometrical structure should be ruled by the curvature scale.
Nevertheless, since, in general, the effect of each component of curvature is not independent, it may hard to distinguish the possibility of conjugate points by the Ricci curvature or the Weyl curvature. 
Our several interesting situations, however, are significantly concerned with only a few leading components, so that the Weyl curvature and the Ricci curvature are not competing. 
Indeed, we prefer the discussion based on such an independence, for example, like Weyl focusing and Ricci focusing\cite{Dyer1981}.
 
In the present article, we see several results for typical situations dominated by the Weyl or Ricci curvature, which would be physically and observationally interested in.
As a first example, in the case of the spacetime with a globally conformally flat region\footnote{Global conformally flatness means the region is described by a single function for the conformal factor.}, we simply have a following theorem about the geometry for the absence of the Weyl curvature.
Since a conformal transformation does not change null geodesics as well known (we will see it concretely in the following subsection), the conjugate point and the wandering null geodesics are not changed by such a global conformal transformation.

%The wandering null geodesic in wandering phase will be possesses
%some typical characteristics similar to the unstable circular orbit of a photon, that is the accumulation of the photon orbits\footnote{It will be shown as the generic structure of such conjugate point in the context of catastrophe theory\cite{FC}.}.
%Then only this wandering phase will relevant for the optical observation
%of the black hole shadow.

%\begin{lemma}
%Without Weyl curvature.
%conjugate point means singularity
%\end{lemma}

\begin{theorem}
A truncated wandering null geodesic is absent in the globally conformally flat region of a spacetime.
\label{th:nogo}
\end{theorem}

Of course, in the case like the Einstein's static universe, since the spacetime is not globally conformally flat, but locally, this theorem does not work and then there are conjugate points. Therefore, null geodesics invade their chronal region.
On the contrary, in fact, the homogeneous region of the Oppenheimer-Snyder black hole\cite{Oppenheimer:1939ue} is covered by a single function of a conformal factor for the global conformally flatness.
That means the truncated future wandering null geodesic is not formed in the region inside of the spherical homogeneous star.
As a straight conclusion, we insist the homogeneous spherical collapsing star excludes such a truncated wandering null geodesic from its inside.
Furthermore, the theorem implies the conjugate point by the Ricci focusing results in the wandering null geodesic in the spherical topology like the Einstein's static universe, which is in the different aspect of that in the Schwarzschild black hole.

Besides, in realistic collapse the assumption of vanishing Wyle curvature and the strict global conformally flatness seems to be too strong.
Rather, we attempt to estimate the exclusion scale of the wandering nature, considering a situation where the Weyl curvature is dominant but not so large and the Ricci curvature is very small to study the Weyl focusing case.
For, the Weyl curvature will be not so large outside of the event horizon and the Ricci curvature would be very small in almost vacuum interstellar space.

\subsection{perturbative Weyl focusing}
In the theorem \ref{th:nogo} we have demonstrated that the conformally flat formation of a black hole is not accompanied by the appearance of truncated future wandering null geodesics.
In general, however, the realistic black hole formation would not be conformally flat, and we expect the Weyl focusing lead the occurrence of the conjugate point like in the Schwarzschild spacetime.  
So, now we consider the possibility the black hole is formed with not so large Weyl curvature.

As in the case of black hole spacetime, most of our interesting cases are vacuum or almost vacuum spacetimes, in which the effect of curvature is dominated by Weyl component. 
Then we will estimate the excluding scale of the truncated wandering null geodesics in such an almost vacuum situation, which is considered to be a region where each curvature components are bounded by a curvature scale $C1<1$, for example, as $ |C_{abcd}C^{abcd}|<C_1^2\ \ \ |R_{ab}R^{ab}|<<C_1^2$.
In such a case, we will eliminate the contribution of Ricci curvature from the deviation equation.

Firstly, we want to distinguish the effect of Ricci curvature $R_{ab}k^ak^b$ from the deviation equation given in the previous section (Eq. (\ref{eq:dev1}) and Eq. (\ref{eq:dev2}))  by an appropriate conformal transformation.
Considering a conformal transformation $\tilde{g}_{ab}=\Omega^2g_{ab}$, while the affine connection $\nabla_a$ changes to $\tilde{\nabla}_b v^a=\nabla v^a+C^a_{bc}v^b,\ \  C^a_{bc}=2\delta^a_{(b}\nabla_{c)}\ln \Omega-g_{bc}g^{ad}\nabla_d \ln\Omega$, the null geodesics are maintained as
\begin{align}
k^b\tilde{\nabla}_bk^a&=k^b\nabla_bk^a+2k^ak^c\nabla_c\ln \Omega-k^bk_bg^{ad}\nabla_d \ln\Omega\\
&=2  \frac{d \ln\Omega}{d s}  k^a,
\end{align}
where $k^a$ is the affine parametrized tangent null vector $(\frac{\partial}{\partial s})^a$.
The affine parametrized tangent null vector $\tilde{k}^a=f(s)k^a=(\frac{\partial }{\partial\tilde{s}})^a$ for $\tilde{g}_{ab}$ geometry, is given by
\begin{align}
f(s)f'(s)+2 f(s)^2  \frac{d \ln\Omega}{d s}  =0,\\
f(s)=-\frac1{c\Omega},\ \ \frac{d\tilde{s}}{d s}=c\Omega^2,
\label{eq:conf}
\end{align}
where $c$ is a constant.

Now we assume that $R_{ab}k^ak^b$ in Eq. (\ref{eq:dev1}) vanishes by this conformal transformation, then the conformal transformation of the Ricci tensor,
\begin{align}
\tilde{R}_{ac}&=R_{ac}-2\nabla_a\nabla_c\ln\Omega-g_{ac}g^{de}\nabla_d\nabla_e\ln\Omega
+2\nabla_a\ln\Omega\nabla_c\ln\Omega-2g_{ac}g^{de}\nabla_d\ln\Omega\nabla_e\ln\Omega\\
\end{align}
requires the conformal factor $\Omega$ satisfies a relation,
\begin{align}
\tilde{R}_{ac}\tilde{k}^a\tilde{k}^c&=\frac1{(c\Omega)^2}\left[k^ak^cR_{ac}-2k^ak^c\nabla_a
\nabla_c\ln\Omega+2\left(k^a\nabla_a\ln\Omega\right)^2\right]\\
&=\frac1{c^2\Omega^4}\left[R_{ac}k^ak^c-2\frac{d^2\ln\Omega}{ds^2}+2\left(\frac{d \ln\Omega}{ds}\right)^2\right]\\
&=0,
\label{eq:tilR}
\end{align}
where we consult $k^ak^c\nabla_a\nabla_c F=k^a\nabla_a(k^c\nabla_cF)-(k^a\nabla_ak^c)\nabla_c F=k^a\nabla_a(\frac{dF}{ds})$.

Here we introduce $\omega(s)$ which satisfies $\omega'=-\omega^2-\frac12[R_{ab}k^ak^b](s)$ along the null geodesic. Then, there exists $\Omega$ such that it satisfies $\omega(s)=-d \ln\Omega/ds$ on the null geodesic congruence.
The smoothness of $\Omega$ will be destroyed by the divergence of $\omega$.
If $\omega(s)$ takes a negative value $\omega_0$ at $s_0$, it diverges within $s-s_0\leq 1/ \omega_0$ provided that $R_{ab}k^ak^b$ is always non-negative. 
This divergence directly means a conjugate point appears by the effect of the Ricci curvature. This, however, is not the case of our almost vacuum situation. So, we will assume the existence of a regular solution for such $\omega(s)$.

By the conformal transformation, the deviation equation of null congruence reduces to 
\begin{align}
\begin{cases}
\dot{\tilde{\theta}}=-\frac12\tilde{\theta}^2-\mathrm{Tr} {\bf \tilde{\boldsymbol{\Sigma}}^2}\\
{\bf \dot{\tilde{\boldsymbol{\Sigma}}}}=-\tilde{\theta}{\bf \tilde{\boldsymbol{\Sigma}}}+{\bf \widehat{C}}\label{matdev},
\end{cases}
\end{align}
where $\tilde{\theta}$ and $({\bf \tilde{\boldsymbol{\Sigma}}})^a_c\equiv \tilde{g}^{ab}\sigma_{bc}$ are for $\tilde{k}^a$ and $\tilde{\nabla}_a$, and $({\bf\widehat{C}})^a_d$ is $\widehat{\tilde{C^a_{bcd}}\tilde{k}^b\tilde{k}^c}$, where the bold face is used to express matrix valued variables.
 $\dot{F }=dF/d\tilde{s}$ is the derivative of $F$ by the conformally transformed affine parameter $\tilde{s}$.

Supposing $\bf\widehat{C}$ is limited by any matrix norm\footnote{A rigorous definition of the norm will not be necessary here. One may define the magnitude of the matrix valued variables by any definite norm of matrices.} $||.||$ as $||\widehat{\bf C}||< C_0<1$, we will consider perturbations by small $\bf \widehat{C}$ expansions.

By the series expansion $F=\sum_i F_i$ of the variables for the order of $\sim C_0^i$, we define
\[
	\tilde{\theta}=\tilde{\theta}_0+\tilde{\theta}_1+O[C_0^2],\ \ \  \tilde{\boldsymbol{\Sigma}}={\bf \tilde{\boldsymbol{\Sigma}}_0}+{\bf\tilde{\boldsymbol{\Sigma}}_1}
+{\bf O}[C_0^2].
\]
The 0-th order background solution is given by
\begin{align}
\tilde{\theta}_0(\tilde{s})=\frac{\theta_{ini}}{\tilde{s}-\tilde{s}_0}, \ \ \ {\bf\tilde{\boldsymbol{\Sigma}}_0}(\tilde{s})=\frac{\boldsymbol{\Sigma}_{\bf ini}}{\tilde{s}-\tilde{s}_0} ,
\end{align}
where $\theta_{ini}$ and $\boldsymbol{\Sigma}_{\bf ini}$ satisfy algebraic equations deduced from the equation (\ref{matdev}) without source term.
This solution corresponds to straight light ray, and of course not divergent with initial conditions $\tilde{\theta}_0>0$, of expanding congruence.
The equations of the 1-st order are given by  
\begin{align}
\dot{\tilde{\theta}}_1&=-\tilde{\theta}_0\tilde{\theta}_1-2\mathrm{Tr}{\bf \tilde{\boldsymbol{\Sigma}}_0}{\bf\tilde{\boldsymbol{\Sigma}}_1}=-\frac{\theta_{ini}}{\tilde{s}-\tilde{s}_0}\tilde{\theta}_1-\frac{2}{\tilde{s}-\tilde{s}_0}\mathrm{Tr}(\boldsymbol{\Sigma}_{\bf ini}{\bf\tilde{\boldsymbol{\Sigma}}_1}),\\
{\dot{\tilde{\boldsymbol{\Sigma}}}_{\bf 1}}&=-{\bf \tilde{\boldsymbol{\Sigma}}_0}\tilde{\theta}_1-\tilde{\theta}_0{\bf\tilde{\boldsymbol{\Sigma}}_1}+\widehat{\bf C}
=-\frac{\boldsymbol{\Sigma}_{\bf ini}}{s-s_0}\tilde{\theta}_1-\frac{{\theta}_0}{\tilde{s}-\tilde{s}_0}{\bf\tilde{\boldsymbol{\Sigma}}_1}+\widehat{\bf C}.
\end{align}

%Then two equations decouple as
%\begin{align}
%\begin{cases}
%(\boldsymbol{\Sigma}_{\bf %ini}\tilde{\theta}_1-\sqrt{2}%{\bf\tilde{\boldsymbol{\Sigma}}_1})^\cdot=-\sqrt{2}\widehat{C} \\
%(\tilde{\theta}_1\mathbb{I}+\sqrt{2}{\bf\tilde{\boldsymbol{\Sigma}}_1})^\cdot=-2\frac{\tilde{\theta}_1+\sqrt{2}\tilde{\sigma}_1}{\tilde{s}-\tilde{s}_0}+\sqrt{2}\widehat{C} 
%\end{cases}
%\end{align}

%Their solution is given by
%\begin{align}
%&\tilde{\theta}_1=-\frac1{\sqrt{2}}\int\widehat{C} ds
%+\frac1{\sqrt{2}}\frac{\int\widehat{C}(\tilde{s}-\tilde{s}_0)^2ds}{(\tilde{s}-\tilde{s}_0)^2}\\
%&\tilde{\sigma}_1=\frac12\int\widehat{C} ds
%+\frac12\frac{\int\widehat{C}(\tilde{s}-\tilde{s}_0)^2ds}{(\tilde{s}-\tilde{s}_0)^2}
%\end{align}
The homogeneous solution of these first order differential equations will be given by $\tilde{\theta}_1=\tilde{\theta}_0/(\tilde{s}-\tilde{s}_0),\ \   {\bf \tilde{\boldsymbol{\Sigma}}_1}={\bf\tilde{\boldsymbol{\Sigma}}_0}/(\tilde{s}-\tilde{s}_0)$.
Then the order of magnitude of the inhomogeneous solutions are of order $\sim ||{\bf\widehat{C}}|||\tilde{s}-\tilde{s}_0|$.
From the initial conditions for 0-th order, $s-s_0$ is positive and this first order solutions are not divergent.
Consequently, as long as the perturbation is consistent, there appears no conjugate points.

%Since the Weyl curvature is bounded by $C_1$, the ratio between $|\theta_1|/|\theta_0|$ or $|{\bf\tilde{\boldsymbol{\Sigma}}_0}|/|{\bf\tilde{\boldsymbol{\Sigma}}_1}|$ never exceed one.

Since the order of magnitude of 0-th order solution is $\sim1/|\tilde{s}-\tilde{s}_0|$ and that of 1-st order solution is $\sim ||{\bf \widehat{C}}|||\tilde{s}-\tilde{s}_0|$.
Within $ |\tilde{s}-\tilde{s}_0|\ll 1/\sqrt{C_0}$, the perturbation is consistent. Therefore, there is no conjugate point for this $\tilde{g}_{ab}$ geometry in such a region.
In the original geometry $g_{ab}$, from Eq. (\ref{eq:conf}) and $\tilde{C}^a_{bcd}\tilde{k}^b\tilde{k}^c=\frac1{c^2\Omega^4}C^a_{bcd} k^bk^c$ that condition turns to   $ |s-s_0|\ll1/\sqrt{C_1}$ when $C^a_{bcd} k^bk^c$ is bounded by $C_1$ as $|C^a_{bcd} k^bk^c|<C_1 <1$.

In this subsection we have argued the existence of the conjugate point, expecting we can treat the geometry around the event horizon by small Weyl curvature expansion.
Around a general black hole, however, the Weyl curvature may not be as small as that around a Schwarzschild black hole (each component of the Weyl curvature in orthogonal basis is order of $2m/r^3$). Then we consider the condition is not general one and adopt the assumption as only a superficial one. 
Under such a difficulty, one may require analyzing it in higher order contributions.

Nevertheless, we consider the more complicated analysis is not so meaningful here because the location of the conjugate points will not be independent of the initial conditions of the null geodesic congruence which is strongly ruled by the circumstances of the optical observations.
Unlike the theorem \ref{th:nogo}, we allude only lightly that, within this scale $\sim 1/\sqrt{|C^a_{bcd}k^bk^c|}$ of affine length, there exists no truncated wandering null geodesic.
Anyway, we see that at the formation of such a massive star with small Weyl curvature dominating, the formation of the truncated wandering null geodesic will be delayed by timescale of this affine length.

\section{relevance to the black hole shadow}

Now we will discuss the relevance of the wandering null geodesics to the black hole shadow.
As well known, the black hole shadow in Schwarzschild spacetime is explained by the photon sphere which is the sphere determined by its circular photon orbits. 
Since we have seen the totally wandering null geodesic coincides with the circular photon orbit in Schwarzschild spacetime, we might expect, in general, the wandering null geodesics are relevant to the black hole shadow.
Here we should be care about that we see the black hole shadow in static spherical black hole spacetime is relevant to the unstable circular orbit.
There, only an unstable circular orbit can collect a photon and form a photon sphere.

Are the wandering null geodesics relevant to the black hole shadow when the black hole is not static?
Here, we attempt to insist that the existence of wandering null geodesics should be significantly related to the contrast of a photon in the field of view for optical observations.
First of all, one may doubt that in dynamical situation the black hole geometry can form such a collection of the photon that is relevant for the optical observation for the black hole, since the position of the circular orbit for a photon is changing.
Though one may tend to feel that the contrast of photons is the result of the photons unstably\cite{Keir2016,Cardoso2014,Cunha2017a} running the same orbit repeatedly from the discussion in Schwarzschild spacetime, we want to demonstrate that actually, in global sense, this unstable aspect corresponds to the accumulation of the null geodesic in a congruence of the null geodesic.

To clarify the relation between the stability and the accumulation, we will require the further analytical consideration for the global configuration of the geodesic congruence. That cannot avoid significant difficulty, since the concept of the stability is usually determined by not global causality but instantaneous kinematics.
In this section, however, in a sense of global causal structure we argue the accumulation and will suggest that any totally wandering null geodesic means that at least one wandering null geodesic with the accumulation exists in asymptotically flat spacetime.
So, the result might be applicable to wider situation if the black hole is sufficiently old.

%%%%%%%%%%%%%%%%%%%%%%%%%%%%%%%%%%%%%%

The relevance of the wandering set would be a subject globally investigated with asymptotic structures of the black hole spacetime.
In the context of global differential structure, such a notion will be realized by the existence of conjugate points.
%For convenience of the discussion, we will artificially consider the number of conjugate points on a null geodesic congruence.
We suppose a situation that there is at least one totally wandering null geodesic through $p_w$---we will generally discuss the black hole shadow incorporating also future (past) wandering null geodesics and event horizon in our forthcoming work\cite{GG3}.
Usually we will consider a congruence\cite{Wald:1984rg} of null geodesics starting from a point $p$ varying the directions of their initial tangent null vectors (or starting from a surface $\cal S$ with orthogonal tangent vectors to it), which is realized as an embedding of the family of geodesics, such that the deviation vector which is the solution of the geodesic deviation equation and the null geodesic congruence construct a coordinate system in order to formulate the deviation equation.
Nevertheless, of course, that would be singular at any conjugate point which is the zero of the deviation vector.

\begin{figure}[hbtp]
\begin{center}
 \includegraphics[height=12cm,clip]{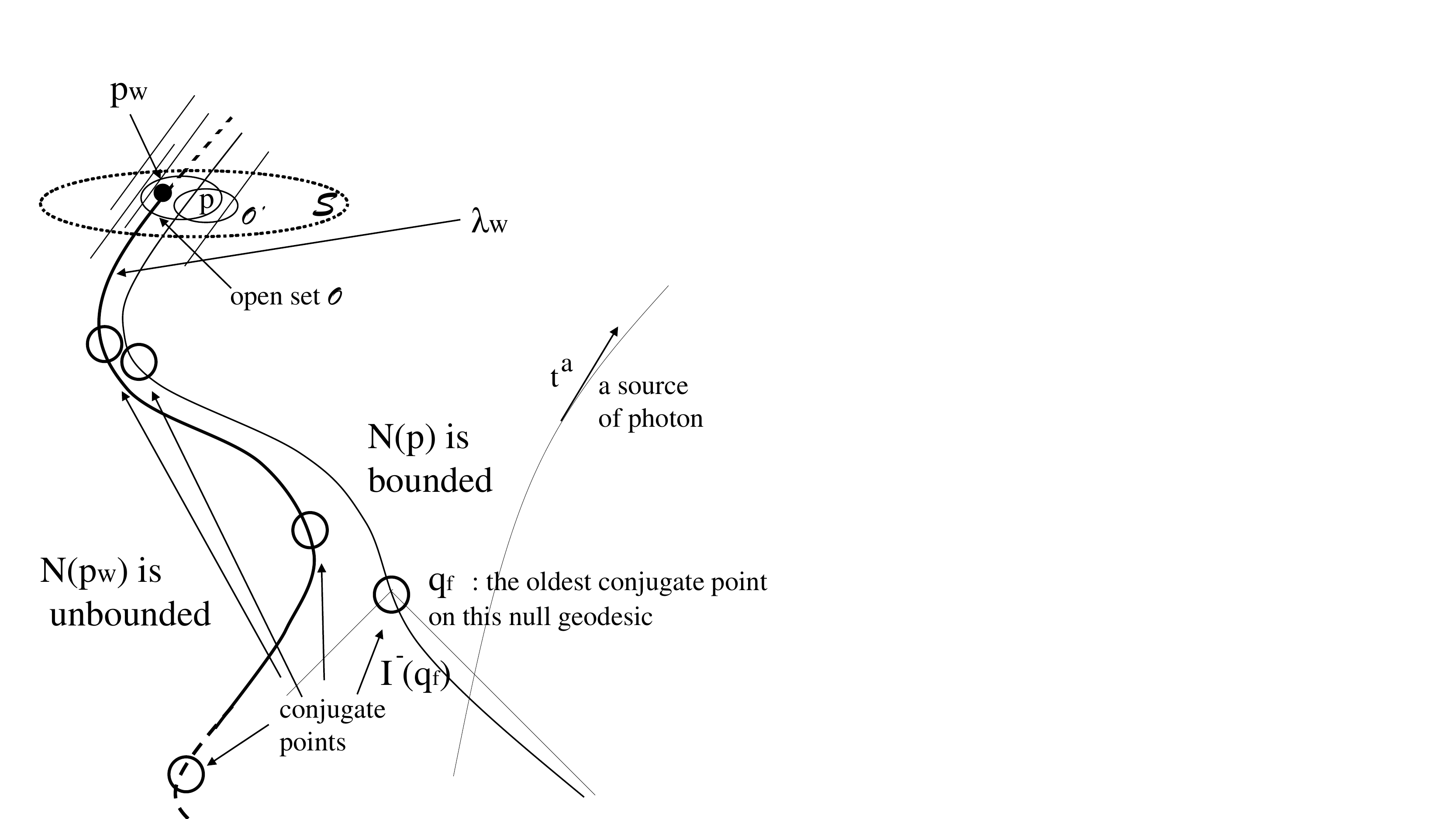}
 \end{center}
 \caption{Singular congruences are considered around a totally wandering null geodesic.
 While there are an infinite number of conjugate points on the totally wandering null geodesic, there are a finite number of conjugate points on a past directed null geodesic starting from $p$. 
The final (oldest) conjugate point $q_f$ determines a past set $I^-(q_f)$ and a segment of the null geodesic from $q_f$ should be a generator of the boundary of $I^-(q_f)$.
 } \label{fig:5}
\end{figure}

Now we are relaxing the concept by extending the geodesic further beyond the singularity of the coordinate and then we allow the singularity of the coordinate as long as each null geodesics is complete.
Also, we relax and generalize the initial condition of the null geodesics as a given smooth null vector field $k^a(p)$ on the initial section ${\cal S}(\ni p)$ which is an open subset of an appropriate spatial 2-surface\footnote{Of course, it is possible to consider the set of null geodesics starting from a point $p$, by considering the initial data set on a surface in the tangent space $T_ {p_w} $ at $p_w$.}, which would be arranged for characteristics of  each observation.
 Then the composition of this `singular' congruence is a set of past null geodesics intersecting an initial section $\cal S$ of the congruence (see Fig. \ref{fig:5}). We denote it as $c[{\cal S}]$.
 The deviation equation will be analytically continued through the conjugate points without a trouble since it is a second order ordinary differential equation.
 
 For convenience of discussion, we artificially consider a number function $N(p)$ on $\cal S$ of the conjugate points along the null geodesic congruence.
Suppose a past directed null geodesic congruence is including a totally wandering null geodesic $\lambda_{wT}$ intersecting $\cal S$ at $p_w$ and the spacetime is geodesically complete in past direction
\footnote{Of course, this is not the case for our cosmology with initial singularity. For such a big bang universe, we will carry out a quantitative investigation, for example the comparison of the magnitude of $N(p)$ with the age of the universe should be significantly discussed.}.
%%%%%%%%%%%%%%%%%%%%%%%%%%%%%%%%%%%%%%%%%%%%
Each geodesic of the congruence $c[{\cal S}]:={\cal S}\times (-\infty,0]$ is defined by the tangent null vector field $k^a$ of the past directed null geodesics and the past directed exponential map $q_p(t)=\exp (k^a t)\circ p, \ \  (p\in {\cal S}, t\in  (-\infty,0])$ (cf.  the embedding of the congruence into the spacetime is not diffeomorphic since there would be conjugate points).
Then we can give any smooth number function of conjugate points $n (p, t) $ such that 
\begin{align}
&N(p)=\max_t\{n(p,t)\},\\
&n(p,t):c[{\cal S}]\mapsto \mathbb{R}, \\
&n(p,t_j(p))=j, \\
& t_{i-1}(p)>t>t_i(p)\ \Rightarrow i-1< n(p,t)< i,
\end{align}
where $i$-th conjugate point $q_i[p]$ of $\lambda_p$ is related to the parameter $t_i$ as
$q_i[p]=q_p(t_i)=\exp(t_i k^a)\circ p$.
Finding such a smooth function $n(p, t)$ on each $\lambda_p$, with the smoothness of the spacetime and the Stone-Weierstrass theorem\cite{Stone1937}, $n(p,t)$ and $N(p)$ can be a smooth function in a domain not containing any $p_w'$ which is an intersection of a totally wandering null geodesic\footnote{Even if two null geodesic with different number of conjugate points is near at a moment, they can be divided into former past region since they will be a generator of different achronal boundary.}.
 %%%%%%%%%%%%%%%%%%%%%%%%%%%%%%%%%%%%%%%%%%%%%%%%%%%
 
 Suppose in any open set $\cal O$ containing $p_w$ include its open subset $\cal O'$ from which no null past geodesic becomes a wandering null geodesic.
 A segment from $q_f$ of such a past directed null geodesic, where $q_f$ is the oldest conjugate point (the final one in the past direction), should be the generator of $\partial I^-(q_f)$ under the global hyperbolicity (as illustrated in Fig.\ref{fig:5}).

On the other hand, this boundary $\partial I^-(q_f)$ is the achronal boundary which is a closed embedded achronal $C^{1-}$ submanifold\cite{Hawking:1973uf}.
We consider a Cauchy surface $\cal C$ and a smooth timelike vector field $t^a$ whose integral curve can be regarded as a model for orbits of luminous sources, since every integral curve of $t^a$ intersects $\cal C$ precisely once. Then homeomorphism $\psi: \partial I^-(q_f)\mapsto {\cal C}$ by the integral curves is given.  Therefore, the image of $\partial I^-(q_f)$ assigned by ${\cal A}=\psi[\partial I^-(q_f)]$ is homeomorphic to the $C^{1-}$ submanifold $\partial I^-(q_f)$. After all, $\cal A$ is a closed and open subset in the topology of $\cal C$, so that $\cal A$ is $\cal C$.
Then all the timelike orbits of the luminous sources intersecting the Cauchy surface $\cal C$ crosses this achronal boundary $\partial I^-(q_f)$ precisely once.

The `sub-congruence' $c[{\cal O'}]$ is also complete and inextendible, and ${\cal C}\backslash \psi[c[{\cal O'}]]$ is not dense in $\cal C$.  Then, with a non-compact Cauchy surface on which there is no bias in the distribution of luminous sources, each null geodesic of the sub-congruence $c[{\cal O'}]$ will cross an infinite (numerous) number of orbits for the luminous sources since the past directed null geodesic is complete and inextendible---though these may be far from astrophysical generality.
 
If $N(p)$ is smooth except for $p_w$, the range of $N(p)$ is very wide in any small domain ${\cal O}\backslash\{p_w\}$ including various positive integer.
Since different positive integer of $N(p)$ means each past directed null generator from $q_f$ on a past directed null geodesic from $p$ is of an achronal boundary for a definitely different past set, with a smooth $n(p,t)$. As the deviation equation is smooth, conjugate points on a same null geodesic are with any separation, they will be destined to entirely different region at far in the past, in general.
Then, the closer $\cal O'$ approaches $p_w$, the wider the range of $N(p)$ becomes and the brighter the light rays are.

On the contrary, for a congruence $c[{\cal O''}]$ of $\cal O''$ from which every null geodesic is the totally wandering null geodesic, at least a class of smooth timelike vector fields ${t'}^a$ exists such that $\psi[c[{\cal O''}]]$ is a compact subset of $\cal C$, which imply there is a strong contrast between $\cal O'$ and $\cal O''$ on the field of view concerning the brightness/darkness around the totally wandering null geodesic. 
Of course, such a designed choice of $t^a$ may significantly affect the result.
We think that should be discussed carefully from the viewpoint of observational reality in time.

Anyway, these ideal discussions logically realize that such totally wandering null geodesic is relevant in optical observation.
For example, that would be corresponding to the fact that, as well known in static spherically symmetric spacetime, the relevance of the circular orbit to the black hole shadow depends on the stability of the null geodesics around the circular orbit of a photon rather than on the aspect of a single photon running on the same circular orbit.

 From the asymptotic flatness, for a sufficiently large $\cal S$ we will say not all null geodesics of the congruence $c[{\cal S}] $ can be totally wandering null geodesics. And then there must be a totally wandering null geodesic $\lambda_{rw}$ from $p_{rw}$ such that any small $\cal O$ containing $p_{rw}$ must include such an open subset $\cal O'$ not containing any intersection of a totally wandering null geodesic.
 Therefore, there should exist at least one such a totally wandering null geodesic $\lambda_{rw}$ relevant to the black hole shadow.
 
   In the case of a spherically symmetric spacetime, this result can be translated as the following.
 To put it shortly, 
The relevance, that is, whether a circular orbit of a photon is unstable or not, is decided by whether an extremal value of the radial potential of null geodesics is maximal or not.
If there is an extremal of the potential function for the radial component of the null geodesic equation, the potential must have any local maximal since it is downwardly convex at a large radius by the asymptotic flatness.
 So, this maximal of a circular orbit will be unstable and is relevant to the black hole shadow of this asymptotically flat spacetime.
  
In order to discuss the dark areas of the photon image in detail, it is not sufficient not to be concerned with the presence of the event horizon as well as totally wandering null geodesics.
 Though that will be complicated in dynamical situation, the cooperation of the event horizon and the wandering null geodesics will become important in our future task.
 
 Of course, the present argument will be applicable to sufficiently old black hole since its conjugate points are sufficiently a lot.
On the contrary, the argument will not be valid for a young black hole since the number of conjugate points, which significantly reflects the contrast of the image of a photon, is sensitive to its geometry. According to the argument in the previous section that would rely on the scale of the Weyl curvature rather than the age of its event horizon.

%%%%%%%%%%%%%%%%%%%%%%%%
The behavior of the number function $n(p,t)$ might directly be determined by the characteristics of the deviation equation.
Then, in principle, the contribution of the curvature tensor is able to be investigated along it.
As shown in the previous section, the Weyl curvature becomes a source of wandering.  So, it is expected that the principal null direction may play any essential role about such a configuration of the wandering null geodesics, and that the null direction would be significantly concerned with the Petrov type classification, of course, which will be related to whether the spacetime is a black hole spacetime or not.
Moreover, as some qualitative aspects of the problem reflects the topology of the wandering set in the space of null geodesics, it may be possible that any topological analysis about the number function $N(p)$ of conjugate points   mathematically rules the relevance of any certain wandering null geodesics.

%For example, if  $1/N()$ were analytic function under a certain condition about geometry, it would conclude that the complement of wandering set is dense in the section and all the wandering null geodesic should be relevant.

 % Especially we may be interested in the case of Ricci flat situation as an ideal case.

%Moreover one may be careful about the accuracy of the observation.
%As it is suggested, for example, the Ricci focusing is resembled by Weyl focusing\cite{Dyer},
% the relevance also could be discussed in a different framework for an averaged larger scale contrast on % %the field of view.
 
On the other hand, you may think that such discussions are all about mathematical things, and in the end it may seem that they cannot make very concrete and powerful claims. However, the other theorem for the existence of wandering null geodesics will be introduced in our forthcoming research, which insure the general existence of the wandering null geodesics approaching to photon sphere. If it is used, the relationship between the dark part and the wandering null geodesics in the imaging of the black hole would be clarified as a more concrete claim.

At present, it is difficult to say that the whole picture of what kind of findings can be obtained as an extension of those discussions. For that purpose, we think that it is necessary to analyze the concrete null geodesic congruence in the black hole spacetime solution.
In fact, it is conceivable that the lensing physics\cite{Bozza2002,Tsukamoto2017,Bozza2003,
Eiroa2002,Petters2003,Iyer2007,Shaikh2019a,Stefanov2010} is related to the accumulation of the null geodesics, from the viewpoint of the photon sphere in the context of geometrical optics. Especially, the critical direction-changing of the light ray possibly agrees with the accumulation of the null geodesics, in the sense of deflection angle which might be major studies of the photon sphere in astrophysics. Since their meanings are almost same in static spacetime (see Fig.\ref{fig1}), their correspondence, in order to make use of our argument, rather should be necessary to be analyzed in dynamical spacetime. We should immediately perform such a concrete calculation of the deflection angle as in those references, even in a strongly dynamical spacetime\cite{AmesThorne,Synge1966,Yoshino2019}\cite{Nitta2011} to discuss its relationship with the null geodesic accumulation. However, due to the difficulty of numerical analysis, we are not currently ready for it. Furthermore, we also think that the analysis of accumulation based on the discussion of catastrophe theory may be more effective, but this is also in the examination stage.

\section{summary and discussions}
In order to give causal concept for general black hole shadows, a wandering set, that is, the set of totally wandering null geodesics, has been introduced, instead of the photon sphere.
A wandering null geodesic is defined as a complete null geodesic accompanied by repetitive conjugate points rather than as a one approaching timelike infinity; especially a totally wandering null geodesic corresponds to a cyclic null geodesic on the photon sphere in Schwarzschild spacetime. 
Moreover, we have considered truncated wandering null geodesics to symbolically discuss its formation concerned with the black hole formation.
  Then we have found that the truncated wandering null geodesic is excluded by curvature radius of the Weyl curvature in Weyl focusing.
  Besides, discussing the relevance to the black hole shadow of the wandering set, we have realized that the essence of the black hole shadow relies on not the stationary cycling of the photon orbits which is the concept only available in a stationary spacetime, but the accumulation of null geodesics near the wandering null geodesic.
 With any past wandering null geodesic, it would be implied that such accumulation will be occur somewhere in past null geodesically complete spacetime.

%We will see there is two typical types of black hole formation characterized by the Weyl curvature.
%Also we can numerically confirm the results in the example of Oppenheimer-Snyder
%\cite{YOSHINO} 
%black hole spacetime and Vaidya black hole spacetime, as they are different in whether the formation of black hole region is accompanied by the truncated wandering set.
%By the present generalization of the photon sphere, we will understand the highly dynamical situation of the black hole shadow, for example the collision of the black hole with the gravitational wave radiation, that has been our first task.
%Furthermore even for the isolated and adiabatically evolving black hole, we would concern the wandering set.

To discuss the observational efficiency, it will be important to confirm the generality and stability of the structure of the wandering set. On the other hand, the causality as well as geometrical optics is easy to fit the catastrophe theory.
We will survey the possible catastrophic structure related to the wandering set.
It is the significant issue whether the wandering set is relevant for optical observation of highly dynamical black holes.
In fact, photons might not be piled up so much if the age of black hole is so young. Nevertheless, as discussed for the Schwarzschild black hole if there is the accumulation of photon orbits, it will be sufficient to consider the usefulness of the wandering set to observe dynamical black holes. Then we shall investigate such accumulation of the photon orbits in general dynamical black holes in the context of the catastrophe theory.

%Here we want to note a technical merit of the wandering set to investigate the black hole shadow.
%To find the photon sphere, in principle, the `ray-tracing' by null geodesic equations would be a realistic and powerful way. Nevertheless, once we know that the relevant ray is a null geodesic accompanied by the conjugate points, the deviation equation will give us further useful information about the kinetics of the photon orbit.
%Moreover as we have found the absence of the wandering set in the conformally flat region, more detailed discussion about the wandering null geodesics may mathematically reveal the appearance of the black hole shadow.

Finally, we discuss the topology of the event horizon.
As discussed in Refs. \cite{Siino:1997ix}\cite{Husa:1999nm}\cite{Cohen:2011cf}\cite{Bohn:2016soe}
\cite{Emparan:2017vyp}, the concept of the dynamical topology of the event horizon is, in a sense, vacant since it is not independent of the definition of timeslicing. Rather, we are interested in the structure of the crease set, which is an acausal arc-wise connected set of the past endpoints of the event horizon\cite{SIINO2011}\cite{Siino2005}\cite{Emparan:2017vyp}.
In fact, any colliding black holes cannot be discriminated from the formation of a prolate black hole in the sense of topology of event horizon.
Especially, a reason making the concept of the event horizon topology unphysical is its variance by the conformal transformation, while we can generate an initial data of spacetime dynamics decoupled from its conformal dynamics\cite{Smarr1978}.
Indeed, Ref. \cite{Husa:1999nm} demonstrated that we have a topological reproduction of conformally transformed spacetime data, where a toroidal black hole can simply be constructed by the conformal transformation from a spheroidal collapse in black hole spacetime.
The present work, however, has suggested that one can distinguish them causally by observing their wandering set, since the existence of the conjugate points never changed by the conformal transformation.

The existence of such a wandering null geodesics suggests another causal concept for black hole spacetime.
At least, in above argument for the result of Ref. \cite{Husa:1999nm}, topological dynamics of the event horizon by the conformal transformation is suggested to be discriminated from the `real' topological dynamics of the event horizon.
The wandering null geodesics might be able to discriminate this degeneracy caused by the change of timeslicing completely.
Since the wandering null geodesic is not included in a null surface, in general, by the black hole shadow it is expected that one can optically observe the topology of the black hole directly from asymptotic region\cite{Nitta2011}.
Furthermore, though it is suggested that the binary black holes can never orbit each other before their coalescence\cite{Siino2009}, the black hole shadow of each black hole (the truncated future wandering null geodesics related to each black hole) may be able to orbit each other.
%In the present work, under certain conditions, the system includes not only null geodesics going to singularity
%but also null geodesics wandering and going to timelike infinity.
Now we expect the existence of the wandering null geodesics will provide another concept to characterize the final state of a heavy star like a black hole. 
We will reveal the various aspects of the wandering null geodesic and wandering set as many authors have done it for the event horizon.
Since the wandering set would reflect the optical observation of such a heavy star, we should bring the investigation of the wandering set forward from both theoretical and observational viewpoints.
In the theoretical study, especially numerical experiments will be used to get important knowledge, as photon orbits should be frequently calculated from the null geodesic equation in ray-tracing.
Nevertheless, if we incorporate the deviation equation, e.g., judging a null geodesic not wandering because of absence of conjugate point by curvature scale, which will make such a numerical analysis further efficient.
The mathematical consideration of their causal structure and the numerical analysis of the photon orbits, are like the two wheels of a cart for studying wandering light paths.

%\section{Acknowledgments}
\begin{acknowledgments}
%We would like to thank Hirotaka Yoshino for fruitful discussions.
 This work has included the relation to the earlier discussion with Daisuke Yoshida and Haruki Terauchi.
\end{acknowledgments}

%---------   References   ---------%
%\input{kumonoito.bbl}
%---------   References   ---------%
\bibliographystyle{unsrt}
\bibliography{sjref}

%\bibitem{FW} in preparation.

% \end{thebibliography}
\end{document}